# Revealing Neurocognitive and Behavioral Patterns by Unsupervised Manifold Learning from Dynamic Brain Data

Zixia Zhou[1], Junyan Liu[1], Wei Emma Wu[1], Ruogu Fang[2], Sheng Liu[1], Qingyue Wei[1,5], Rui Yan[1,5], Yi Guo[3], Qian Tao[4], Yuanyuan Wang[3], Md Tauhidul Islam[1*], and Lei Xing[1,5,6*]

[1] Department of Radiation Oncology, Stanford University, Stanford, California, 94305, USA
[2] J. Crayton Pruitt Family Department of Biomedical Engineering, University of Florida, Gainesville, FL. 32611, USA
[3] Department of Electronic Engineering, School of Information Science and Technology, Fudan University, Shanghai 200433, China
[4] Department of Imaging Physics, Delft University of Technology, Lorentzweg 1, 2628 CJ Delf, the Netherlands
[5] Institute of Computational and Mathematical Engineering (ICME), Stanford University, Stanford, California, 94305, USA
[6] Department of Electrical Engineering, Stanford University, Stanford, California, 94305, USA

## Abstract

Dynamic brain data, teeming with biological and functional insights, are becoming increasingly accessible through advanced measurements, providing a gateway to understanding the inner workings of the brain in living subjects. However, the vast size and intricate complexity of the data also pose a daunting challenge in reliably extracting meaningful information across various data sources. This paper introduces a generalizable unsupervised deep manifold learning for exploration of neurocognitive and behavioral patterns. Unlike existing methods that extract patterns directly from the input data as in the existing methods, the proposed Brain-dynamic Convolutional-Network-based Embedding (BCNE) seeks to capture the brain-state trajectories by deciphering the temporospatial correlations within the data and subsequently applying manifold learning to this correlative representation. The performance of BCNE is showcased through the analysis of several important dynamic brain datasets. The results, both visual and quantitative, reveal a diverse array of intriguing and interpretable patterns. BCNE effectively delineates scene transitions, underscores the involvement of different brain regions in memory and narrative processing, distinguishes various stages of dynamic learning processes, and identifies differences between active and passive behaviors. BCNE provides an effective tool for exploring general neuroscience inquiries or individual-specific patterns.

## Introduction

The brain is a highly complex and dynamic organ, continuously adapting and responding to both external stimuli and internal processes. Recent advances in neuroimaging and electrophysiological techniques, from non-invasive whole-brain methods like functional Magnetic Resonance Imaging (fMRI)[1] and Electroencephalography (EEG)[2] to invasive intracranial recordings with implanted electrodes in non-human primates, have enabled the study of neural activity across diverse spatial and temporal scales. These diverse data sources have provided effective means for illuminating the neural bases of cognitive and behavioral processes, such as problem solving, learning, decision making, and consciousness[3–5].

Dynamic brain data often involves extended, non-periodic cognitive and behavioral processes, further increasing the complexity of the data. When considering the entire time series, the dynamic response of the brain across different recording sites is typically depicted as a matrix, where rows and columns correspond to spatial and temporal variables, respectively. The inherent complexity of the data with intertwined temporospatial relationships, compounded by the sparse and noisy nature of the data, poses a formidable challenge in cognitive and behavioral pattern discovery and analysis of brain behavior[6–9]. Computationally, dimensionality reduction and visualization techniques are often employed to project high-dimensional (HD) vectors representing the brain's spatial responses at given time points into lower dimensions sequentially, obtaining temporal brain state trajectories (Fig.1**a**). Common dimensionality reduction methods (e.g., UMAP[10], t-SNE[11], and PHATE[12]) treat brain activity as a series of instantaneous snapshots, which often yield fragmented and spurious embeddings. Recent efforts have improved dynamic brain data visualization by explicitly incorporating temporal continuity, notably T-PHATE[13], which utilizes temporal context for trajectory denoising, and CEBRA[14], which employs contrastive learning for meaningful structure identification from neural recordings. Our proposed Brain-dynamic Convolutional-Network-based Embedding (BCNE) approach advances dynamic brain data analysis by combining the strength of T-PHATE in temporal signal processing and the deterministic advantages of deep neural network-based techniques such as CEBRA. Specifically, BCNE explicitly denoises temporal signals within each spatial channel using an autocorrelation-based affinity matrix, as inspired by T-PHATE. To incorporate the interactions among recording channels, an image representation is introduced for each time point, which maps the responses of the spatial channels at that time point to image pixels in a manner that the intrinsic inter-channel relationships are uniquely encoded through the image contextual pattern (see Fig. 1b and *Methods* for details). These structured images are then processed by a convolutional neural network (CNN)-based framework, providing unsupervised dimensionality reduction of temporospatial data by progressively minimizing the Kullback-Leibler (KL) divergence between the pairwise similarity distributions in their HD and low-dimensional (LD) representation spaces (Fig. 1c). BCNE thus generates structured 2D embeddings that effectively integrate temporal dynamics with spatial relationships, significantly improving the detection and interpretation of subtle cognitive and behavioral patterns.

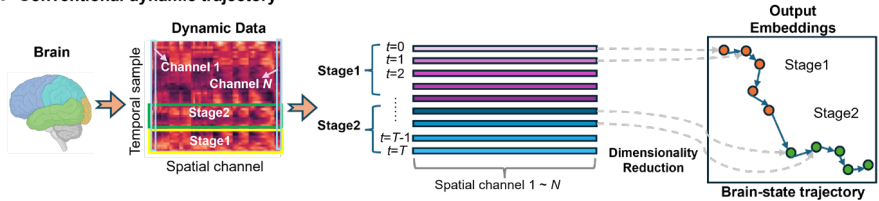

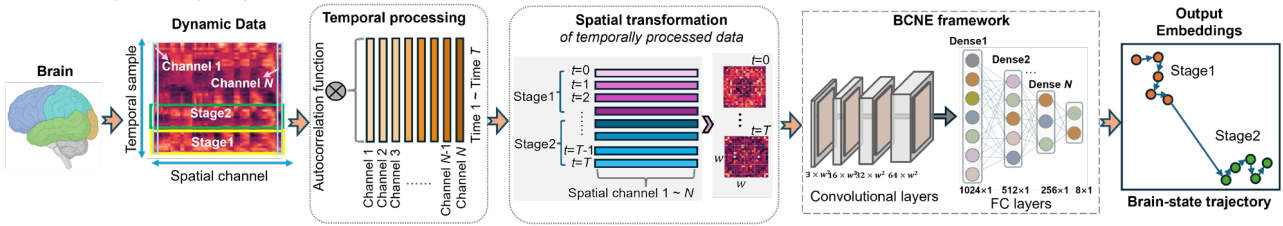

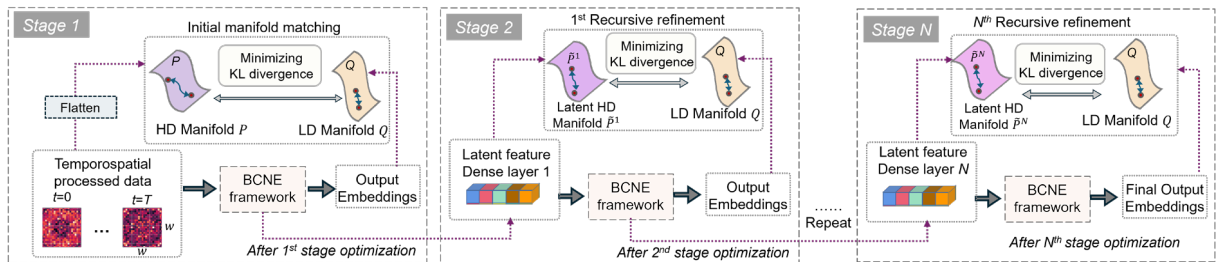

**Figure 1. Schematic of the brain-dynamic trajectory visualization pipeline.** a, Workflow of a conventional dimensionality reduction approach for understanding the evolvement of brain responses during behavioral transitions. b, Workflow of the proposed BCNE: Original dynamic brain data (dimension $N \times T$) are first processed through temporal and spatial modules to produce structured images (dimension $N \times w \times w$). These structured images are then fed into the BCNE framework, which progressively optimizes the brain-state trajectory visualization by minimizing the KL divergence at each stage. c. Detailed illustration of the progressive BCNE network optimization process. At each recursion step $r$ ($r=1, …, R$), latent features obtained from dense layer r become the input data representation for the subsequent stage. At each stage, the BCNE network is optimized by minimizing the KL divergence between the reference distribution (from the initial embedding $P$ to the progressively refined embedding $\tilde{P}^R$) and the current embedding distribution $Q$. This process is repeated, with each stage incorporating a larger temporal or spatial receptive field, until the final stage $R$ is reached.

Specifically, we introduce a principled and broadly applicable deep-learning-based visualization framework, termed BCNE, designed for a wide range of dynamic brain data (Fig. 1 **b** and **c**). BCNE leverages both temporal continuity and the correlational structure among different recording channels, such as fMRI voxels and electrophysiological electrode sites, to uncover meaningful trajectories that reflect underlying cognitive or behavioral states. Although it is well established that interactions across distinct brain states and recording locations are crucial for characterizing cognitive function, these interactions have seldom been fully harnessed for deeper pattern discovery. In BCNE, we begin by constructing a temporospatial representation of the multivariate time series data that captures both temporal dependencies and spatial correlations. From this representation, we apply an unsupervised convolutional framework to learn deterministic mappings that yield structured, interpretable trajectories. To further enhance our ability to uncover subtle patterns, we implement a recursive, manifold-based optimization strategy. This iterative refinement process progressively incorporates deeper-level constraints from the latent representation[15–18], enhancing the model's fidelity and enabling the extraction of more nuanced trajectories that reveal subtle underlying brain activity. It is important to emphasize that BCNE's deterministic nature produces consistent trajectories for identical stimuli under similar conditions, supporting reproducibility, identifying individualized patterns of brain activity, and offering a robust framework for analyzing dynamic brain functions. By capturing intricate temporal dependencies and inter-site correlations, BCNE yields interpretable trajectories that elucidate the structure of cognitive and behavioral processes without the need for additional supervision.

The proposed BCNE technique is showcased using several well-established datasets sourced from diverse origins, including the Sherlock fMRI BOLD dataset, the Rat hippocampus dataset, and the Macaque dataset. Across all datasets, BCNE consistently outperforms traditional HD data interpretation tools in accurately delineating cognitive stages within dynamic learning processes, offering versatility and noteworthy insights into brain dynamics that are unattainable with existing techniques. Specifically, BCNE: 1) closely tracks brain trajectory changes in response to shifting scenes during movie watching; 2) robustly reveals varying degrees of task-related associations across multiple regions of interest (ROIs) during the same behavioral process; 3) effectively distinguishes different learning stages, with brain trajectories gradually diverging between early novice and advanced proficient stages; 4) demonstrates strong underlying information preservation capability, as evidenced by the high correlation between embedding space and the subject's physical movement in geodesic coordinates; and 5) accurately discriminates between active and passive behavior modes by generating distinct embedding patterns. Overall, these findings position BCNE as a powerful tool for exploring general neuroscience inquiries.

## Results

BCNE is a deterministic deep learning framework for visualizing dynamic brain data by jointly modeling temporal dependencies and spatial-channel interactions. Its methodological pipeline unfolds as follows, with details given in *Methods* section. (i) Temporal processing (Fig. 1**b**): To emphasize salient temporal dependencies while suppressing high-frequency noise, we calculate the autocorrelation function for each recording channel, determine the lag threshold at which correlations sharply decline, and generate a lag-weighted average signal representation. (ii) Generation of structured image representation of spatial-channel responses (Fig. 1**b**): To incorporate the interactions among recording channels, we introduce a schematically meaningful image representation for each time point, which maps the channel responses at that time point to image pixels in a manner that the inter-channel relationships are uniquely encoded in the image contextual pattern. Specifically, we compute pairwise correlations as the inverse covariance of the temporally processed signals, then find the optimal transformation to map the channel responses at that time point to a 2D image by minimizing the Gromov–Wasserstein discrepancy between the interaction matrix and pixelated grid-distance matrix[19]. (iii) Feature Extraction from the generated images via BCNE (Fig. 1**b** and 1**c**): The interactive features encoded in the image patterns of the data are extracted through a convolutional neural network (CNN), termed BCNE, to generate a LD embedding of the imaging data by minimizing the KL divergence between the joint-probability matrices of the HD input data and their LD embeddings (the dashed box in the left of Fig. 1**c**). (iv) Recursive refinement of the outcome (Fig. 1**c**): To further improve the resultant trajectory of the brain dynamic data, the initial HD probability matrices used for network optimization are replaced by feature vectors extracted from the first dense layer of BCNE (Fig. 1**c**). Subsequently, BCNE fine-tunes the KL divergence between the joint probability matrices, yielding a refined LD embedding. This procedure continues in two to three iterations by using the feature vectors from progressively deeper dense layers of BCNE, providing a well fine-tuned temporal trajectory representation of the dynamic brain data.

### BCNE Effectively Reveals Scene-Dependent Brain Trajectories and Task-Related Associations Across Regions from fMRI BOLD data

We explored the Sherlock dataset using various embedding approaches to illustrate the potential of BCNE for understanding fMRI BOLD signals during BBC *Sherlock* movie viewing (Fig. 2**a**).

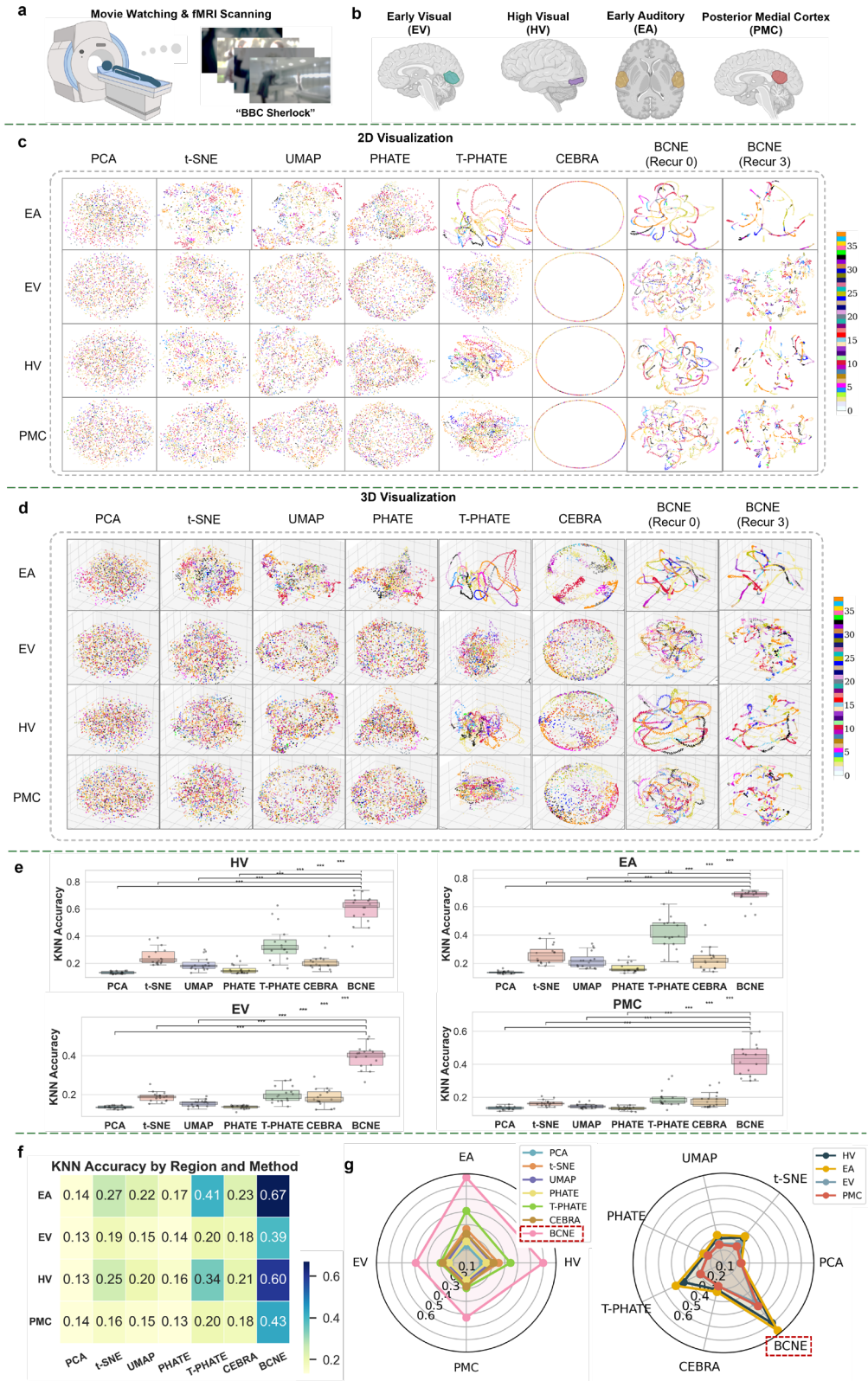

**Figure 2. Visualization and analysis of the Sherlock fMRI dataset.** All results are averaged over three random seeds, with the mean value used as the final evaluation criterion. **a**, Schematic illustration of the fMRI BOLD signal acquisition process during viewing of the BBC Sherlock movie, involving 16 subjects. Movie-related images within the figure have been intentionally obscured to comply with copyright restrictions. **b**, ROIs selected for embedding and visualization. **c**, Comparative 2D visualizations generated by PCA, t-SNE, UMAP, PHATE, T-PHATE, CEBRA, BCNE (Recur0), and BCNE (Recur3) for four ROIs: EV, HV, EA, and PMC, with data colored by movie scene. **d**, Comparative 3D visualizations generated by the same methods and colored by movie scene. **e–g**, Boxplots, heatmap, and radar charts showing the average k-nearest neighbor (KNN) classifier accuracy for the movie scene classification (39 categories) task based on 2D embeddings generated by PCA, t-SNE, UMAP, PHATE, T-PHATE, CEBRA, and BCNE (Recur3) across the four ROIs. For all boxplots, the center line indicates the median, box limits denote the interquartile range (IQR; 25th–75th percentiles), whiskers indicate the full data range, and individual points represent each subject ($n = 16$ per group). The shaded rectangle within each box highlights the dense region (40th–60th percentiles), using a darkened box color to emphasize central tendency. Overlayed dot plots display individual subject values. The unit of analysis is a single human subject; all $n$ values refer to distinct subjects. No technical replicates were used. Statistical comparisons between groups were performed using two-sided independent $t$-tests unless otherwise indicated. *$P < 0.05$, **$P < 0.01$, ***$P < 0.001$. For all pairwise comparisons, the $t$-statistic, degrees of freedom, and exact two-sided $P$ values are provided in the Source Data. Only significant comparisons are annotated in the plots. Box colors correspond to the methods indicated on the x-axis. Unless otherwise specified, the same statistical testing and box plot conventions are applied in Fig. 3 and Fig. 4.

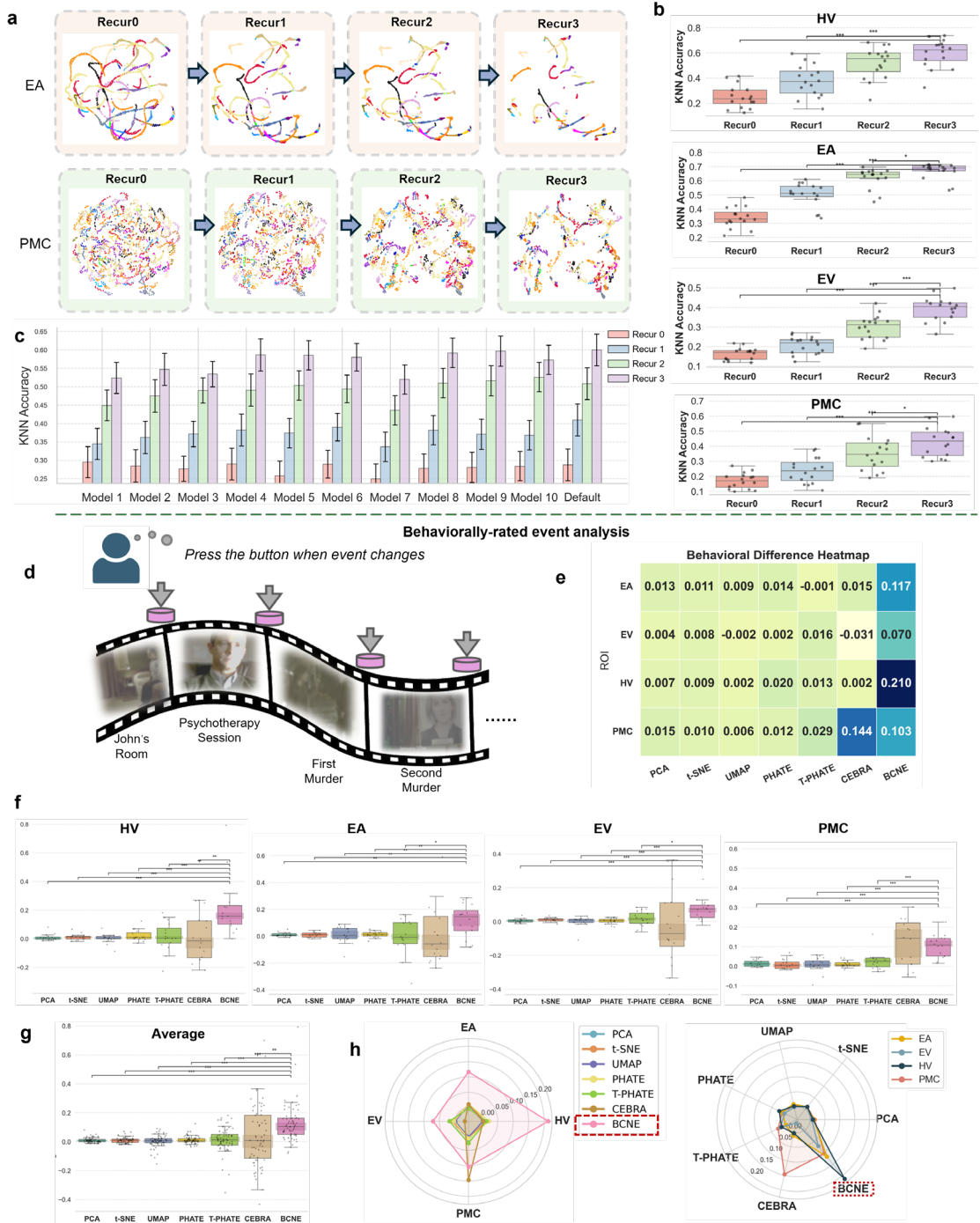

**Figure 3. Evaluation of the BCNE framework and behaviorally-rated event analysis of the Sherlock fMRI dataset. a**, Examples of 2D visualizations generated by BCNE with varying recursion depths (Recur 0–3). Colors correspond to the same scheme defined in Fig.2c. **b**, Boxplots showing KNN classifier accuracy calculated from 2D embeddings generated by BCNE at different recursion depths ($n$ = 16 per group). **c**, Description of the process by which behavioral raters generated event segmentation labels. **d–g**, Heatmaps, boxplots, and radar charts comparing results of behaviorally-rated event analyses across 2D embeddings generated by PCA, t-SNE, UMAP, PHATE, T-PHATE, CEBRA, and BCNE for the four ROIs.

We focused on four ROIs: the early visual cortex (EV), high visual cortex (HV), early auditory cortex (EA), and posterior medial cortex (PMC) (Fig. 2**b**). These ROIs were selected based on prior studies[13,20,21] showing that they generate robust and reliable signals under naturalistic audiovisual stimulation. Each region demonstrated distinct activity levels during the task. Figs. 2**c** and 2**d** illustrate visualizations produced by BCNE and several standard dimensionality-reduction techniques, annotated by 39 movie-scene-related categories. Traditional methods (PCA, t-SNE, UMAP, PHATE) offer limited clarity in capturing temporal and contextual aspects of brain activity.

By incorporating temporal structure, T-PHATE shows noticeable differences among the four regions: EV and PMC exhibit comparatively less variability in perception and comprehension processes, whereas HV and EA reveal stronger trajectories. However, T-PHATE's performance is sometimes unstable (Figs. S1 and S2), struggling to capture detailed temporal dynamics consistently. For CEBRA, we specifically employed CEBRA-Time (see *Methods*), finding that while it can produce richer results in 3D, its 2D embeddings fail to convey as much detail. In contrast, BCNE excels at embedding dynamic signals with stable and interpretable patterns. As shown in Figs. 2**c** and 2**d**, BCNE sharpens the visualization, highlighting intricate morphological trajectories within each ROI more effectively than other methods. Notably, BCNE shows the EV and PMC embeddings to be more dispersed, while the EA and HV embeddings form more cohesive, continuous shapes with clear temporal progression. We next evaluated downstream classification performance by assigning the dimension-reduced embeddings to 39 scene categories using 10-fold cross-validation and a KNN classifier (Figs. 2**e–g**). BCNE achieves the highest accuracy across all four ROIs, improving on T-PHATE by 81.74% (averaged across subjects and ROIs).

Further, Fig. 3**a** and 3**b** show that deeper latent representations from BCNE reveal more distinct cognitive and behavioral transitions, underscoring its ability to capture scene changes in cinematic perception and comprehension. We also examined whether BCNE's performance is sensitive to model structure by comparing ten different architectural variants (differing in the number and width of both convolutional and fully connected layers). Our results indicate that BCNE is robust to these design choices (Supplementary Figs. S5, S6 and S11), and our default architecture achieves the best overall performance with acceptable computational complexity (based on FLOPs comparisons, see Table S1). Finally, we performed behaviorally rated event-boundary analyses to assess how well each method's embeddings capture meaningful cognitive transitions[13,22,23]. This analysis measures alignment between the embeddings and event boundaries identified by an independent behavioral rater (see *Methods*). As shown in Figs. 3**f-h**, across 16 participants and four ROIs, BCNE outperforms PCA, t-SNE, UMAP, PHATE, T-PHATE, and CEBRA in capturing these transitions. Although CEBRA slightly exceeds BCNE in PMC, its variance is notably larger, and BCNE remains superior in HV, EA, and EV. Overall, these results confirm that BCNE preserves both local (intra-event) coherence and interpretable across-event transitions, offering a comprehensive perspective on cognitive trajectories.

## BCNE Uncovers Learning Stage Transitions from Hippocampal Dynamics

We applied BCNE to a rat hippocampus dataset to reveal the neural underpinnings of behavioral learning. This dataset captures spike activity in the hippocampus while rats traverse a linear track from one end to the other (Fig. 4**a**). Each traversal proceeds without mid-course directional changes, facilitated by reward-based training (water feeding), which encourages the animals to identify and follow an optimal path. This setup thus encapsulates the learning—or memorization—of an efficient navigational strategy.

We first examined how increasing the recursive depth in BCNE influences the clarity of identified learning stages. As shown in Fig. 4**b**, deeper recursion (Recur 0 → Recur 3) more clearly demarcates distinct temporal phases in the neural activity. To quantify these differences, we divided the navigational data into $N$ discrete learning stages for an $N$-class classification task, using a KNN classifier to assess the embeddings. With BCNE, accuracy rose substantially from 0.248 (Recur 0) to 0.470 (Recur 3), a nearly 100% relative improvement, thus underscoring BCNE's ability to uncover meaningful stage-specific structure in hippocampal recordings.

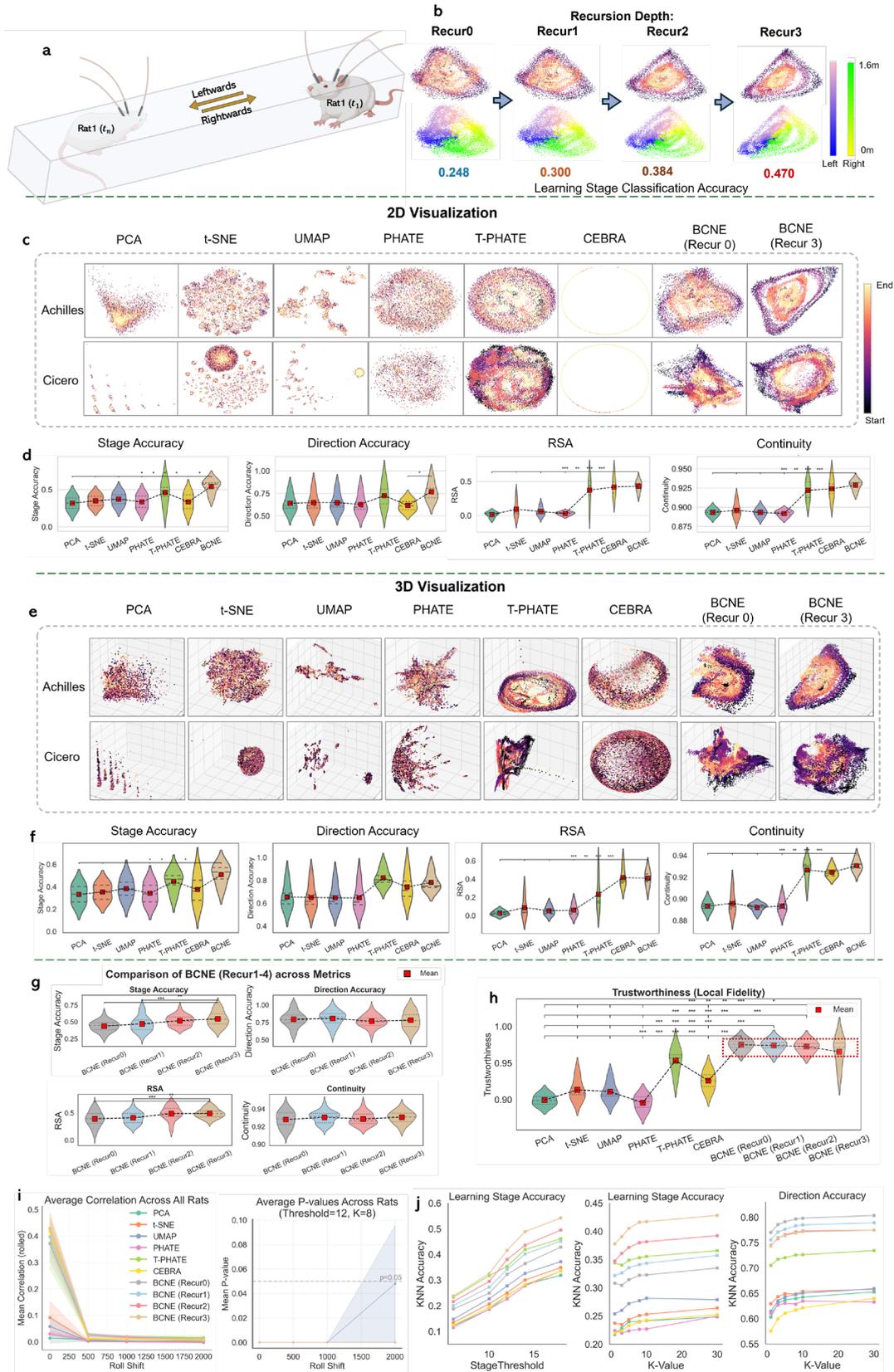

**Figure 4. Results of the Rat Hippocampus dataset.** All results are averaged over three random seeds, with the mean value used as the final evaluation criterion. **a,** Schematic illustration of the rat hippocampus dataset acquisition process (created with BioRender.com). Electrophysiological data were collected while four rats traversed a 1.6 m linear track in either a "leftward" or "rightward" direction. **b,** Example of 2D visualizations generated by

BCNE with varying recursion depths, with classification accuracy for learning stages indicated below each visualization. The upper visualization is colored by learning stage, while the lower visualization uses two distinct color bars to differentiate movement positions for leftward and rightward motions. **c,** 2D visualizations for two randomly selected rats (named *Achilles* and *Cicero*) generated by PCA, t-SNE, UMAP, PHATE, T-PHATE, CEBRA, and BCNE (Recur 0 and Recur 3). **d,** Violin plots of stage accuracy, direction accuracy, RSA, and continuity for the 2D embeddings. **e,f,** Corresponding 3D visualizations and violin plots for the 3D embeddings as described in c and d. **g,** Violin plots showing stage accuracy, direction accuracy, RSA, and continuity, calculated using 2D BCNE with varying recursion depths (Recur 0–3). **h,** Trustworthiness comparison calculated by 2D embeddings across all methods. **i,** Average correlation analysis using roll-shift testing calculated by 2D embeddings across all rats. Correlation values and *p*-values are plotted. **j,** Analysis of learning stage accuracy calculated by 2D embeddings across different stage thresholds and the influence of *k*-values on KNN classifier performance. Violin plots show data from $n = 4$ biological replicates per group (four individual rats). No adjustments for multiple comparisons were made. For panels **i** and **j**, line colors correspond to the methods indicated in the boxplots and are defined consistently across all panels.

We next compared BCNE against PCA, t-SNE, UMAP, PHATE, T-PHATE, and CEBRA, using both 2D and 3D embeddings (Figs. 4**c,e**). For example, in Rat *Achilles*, PCA yields only minor temporal separation (with initial time points situated at the periphery and later points internally), whereas Rat *Cicero*'s PCA results produce more dispersed clusters. Both t-SNE and UMAP largely yield dispersed points without intuitive temporal continuity, although t-SNE sometimes reflects temporal changes. PHATE typically forms circular scatter distributions with limited dynamic information. T-PHATE captures some temporal trajectories but can be less stable across different rats. Meanwhile, CEBRA in 2D appears as an indistinct circular arrangement, although its 3D version does improve clarity somewhat— yet the resulting point clouds remain too scattered for clear trajectory interpretation. In contrast, BCNE consistently captures dynamic behavioral variations in both 2D and 3D visualizations, forming well-defined manifolds that reflect continuous trajectories, demonstrating robustness and stability. Extended evaluations on additional rats (Fig. S8) confirm these findings: BCNE yields the most robust visualizations, extracting meaningful patterns that are otherwise elusive.

In terms of quantitative comparisons (Figs. 3**d,f**), BCNE achieves the highest classification accuracy for learning-stage identification across all tested rats. Additional metrics, including direction accuracy, representational similarity analysis (RSA) with respect to spatial position labels, and embedding continuity, were also examined. Among the 2D results, BCNE consistently outperforms PCA, t-SNE, UMAP, PHATE, T-PHATE, and CEBRA across these measures. For the 3D results, T-PHATE achieves slightly higher direction accuracy, and CEBRA slightly outperforms in RSA. However, BCNE excels across all other indices, demonstrating the overall best performance. We also performed a systematic evaluation of the recursive optimization strategy in the rat dataset (Fig. 4**g**). Although deeper recursion in BCNE generally improves learning-stage differentiation, other performance indices remain relatively stable. To assess local fidelity, we computed trustworthiness (Figs. 4**h-j**). BCNE embeddings (across Recur 0–3) yield higher trustworthiness than all competing methods, confirming stronger preservation of local neighborhood structure. As recursion depth increases, the utilization of latent features in BCNE optimization enables the capture of more global structure but may sacrifice some fine details, resulting in a slight reduction in trustworthiness. A roll-shift analysis further validates these embeddings by showing a marked drop in correlation with large time shifts, indicating that all methods are sensitive to temporal misalignment. P-values for all methods except UMAP remain near zero across shifts, emphasizing their statistical reliability; UMAP shows an increase in p-values with larger roll shifts, indicating reduced significance. Finally, we investigated the impact of varying the learning-stage threshold (e.g., by setting different reward frequencies as the stage division criteria: 6, 10, 12, 14, 18) and *k*-values in the KNN classifier on performance. Across all conditions, BCNE continues to outperform other methods, robustly delineating distinct learning stages and capturing how the rat's navigational strategy evolves with experience. Regardless of the KNN neighborhood size, BCNE consistently achieves the highest accuracy among all compared methods, demonstrating its superior ability to identify and track dynamic hippocampal processes.

### BCNE Discriminates Between Active and Passive Behaviors Through Robust Trajectory Embeddings

We next evaluated BCNE on a macaque sensorimotor dataset in which a macaque performed center-out arm movements along eight distinct directions, under both active and passive conditions (Fig. 5**a**). To visualize these data, we first averaged trials sharing the same target angle, then generated 2D and 3D embeddings separately for the active and passive conditions (Fig. 5**b**). Notably, BCNE was the only method that clearly differentiated between active and passive states in the resulting LD space, whereas PCA, t-SNE, UMAP, PHATE, and T-PHATE all produced similar or overlapping embeddings for the two modes. In previous study[24,25], the experimental results have shown that neurons in Area 2 represent limb states differently depending on the movement type and that the encoding of passive movements differs systematically from active reaches. This aligns with our observations: the BCNE-derived embeddings produce distinct trajectory shapes under active and passive conditions, reflecting differences in how neural signals vary in response to movement type. When comparing the trajectories generated by BCNE from brain data, we observe that active movements produce more linear paths, reflecting a direct reach toward a specific angle, while passive movements result in more curved or circular patterns, aligning closely with the actual motion trails.

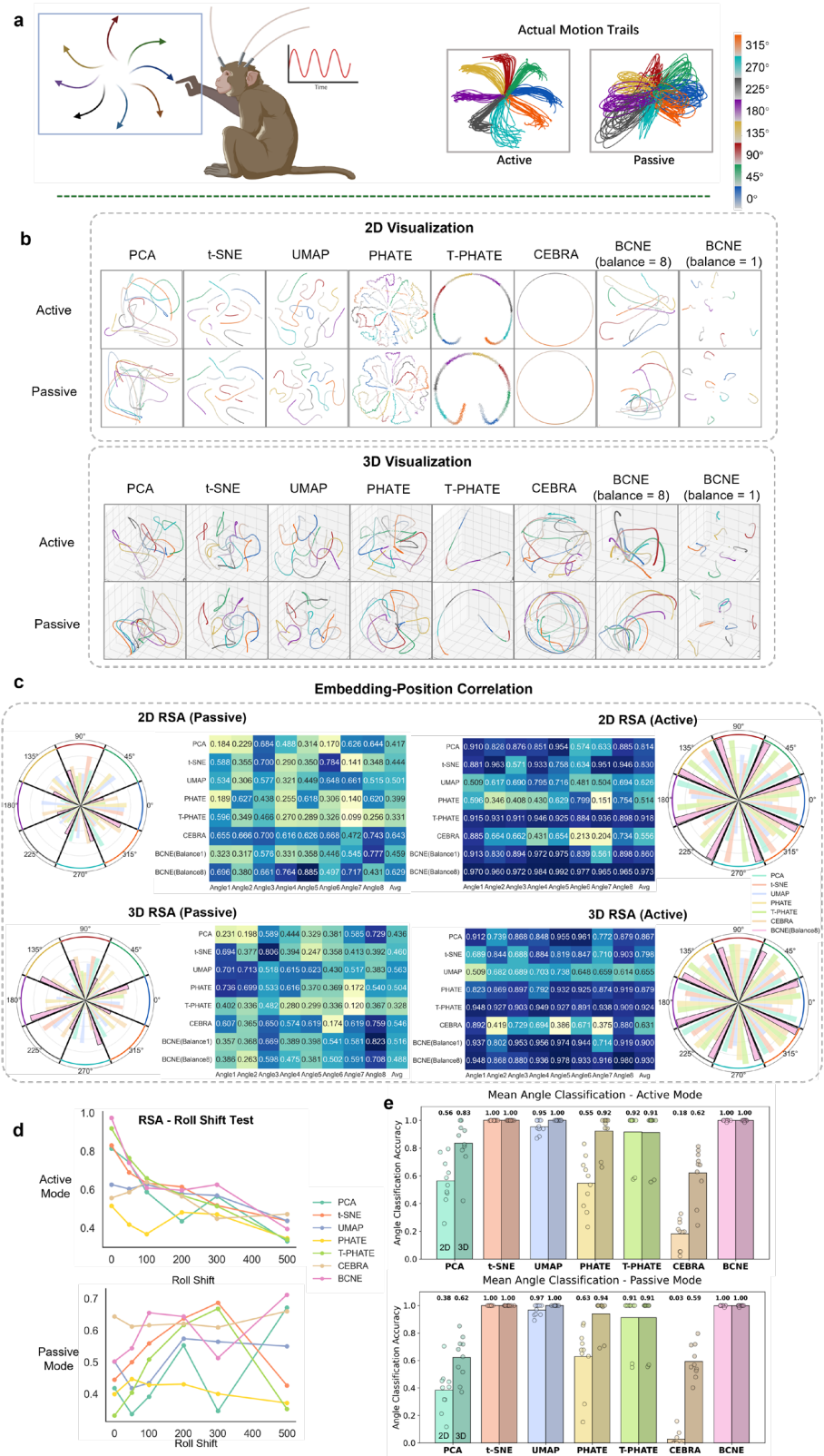

**Figure 5. Results of the Macaque dataset.** All results are averaged over three random seeds, with the mean value used as the final evaluation criterion. **a,** Illustration of the Macaque dataset acquisition process. Data were collected as a macaque performed active or passive movements in eight directions. The actual position of the macaque's arm was recorded and displayed. **b,** 2D and 3D visualizations of active and passive modes, generated by PCA, t-SNE, UMAP, PHATE, T-PHATE, CEBRA, and BCNE. The color bar transitions gradually from light gray to vibrant colors, representing the progression of time during directional reaching. **c,** Circular plots and heatmaps showing the RSA between actual positions for each angle and the embeddings generated by BCNE and all compared methods. **d,** Line charts for roll shift testing calculated by 2D embeddings. **e,** Bar plots showing the reaching angle classification accuracy based on KNN classifiers, using 2D embeddings generated by different methods, for both active and passive modes.

To evaluate the alignment between neural embeddings and actual limb positions, we performed RSA and roll shift test across all eight reach directions (Fig. 5**c** and **d**). The results were visualized using a heatmap that compared active and passive RSA scores across different embedding methods. As shown, RSA values were generally higher in the active mode, indicating a more explicit encoding of position-related information during voluntary movements. Among the methods, BCNE achieved the highest RSA scores in the active mode, with averages of 0.973 (2D) and 0.930 (3D). T-PHATE ranked second, with correlation values of 0.918 (2D) and 0.924 (3D). By contrast, RSA scores in the passive mode were largely unaffected by increasing roll shifts (Fig. 5**d**), implying that fewer task-related signals are present during passive movements and consequently that limb position is less distinctly represented in the neural embeddings.

Lastly, to quantify how well each embedding preserves angle-specific information, we next applied a KNN classifier to predict one of eight reaching directions under both active and passive conditions (Fig. 5e). For this eight-class task, chance performance is only 12.5%. Using both two- and three-dimensional embeddings, t-SNE and BCNE achieved near-perfect accuracy ($\approx$100%), while UMAP and T-PHATE also performed strongly, exceeding 90% accuracy. In contrast, the remaining methods yielded substantially lower scores. These results demonstrate that BCNE not only visually segregates distinct movement types but also retains precise, angle-specific neural dynamics that support highly reliable decoding.

## Discussion

Dynamic brain data analysis is a fundamental task in neuroscience and plays a crucial role in understanding the intricate mechanisms underlying brain activities during cognitive and behavioral processes. By extracting meaningful patterns from multivariate HD data, BCNE provides researchers and clinicians with a valuable tool to gain critical insights into brain function and behavior. In essence, the BCNE leverages the projections of temporospatial correlation to reconfigure the input data into a more conducive representation for pattern discovery. As evident from Figs. 4, 5, BCNE markedly enhances the representational similarity of the embeddings and the continuous positions of subject movement. To evaluate the effectiveness of each component within BCNE, extensive ablation studies were conducted (see Figs. S3, S8, S9 and S10), including versions of BCNE without spatial correlation projection, with 1D autoencoder (AE)-based spatial representation, with 2D AE-based spatial representation, and with an autocorrelation-based spatial representation. These studies clearly demonstrated the superior performance of our BCNE design. We noticed that the autocorrelation-based spatial representation resulted in unstable outcomes, causing gradient vanishing or explosion[26] in network training for some cases. Moreover, the information quantity and discernibility of the projected data generated by the autocorrelation-based spatial representation are inferior to those achieved with our BCNE (see Fig. S9). Temporal correlation proved critical for complex sequences. Within the BCNE pipeline, temporal smoothness is achieved by the autocorrelation-based weighting matrix; no other components impose proximity constraints on adjacent time points. The omission of this temporal smoothing in analyses of the Sherlock fMRI dataset resulted in fragmented embeddings with poor continuity. Conversely, the Macaque dataset showed smaller performance improvements, reflecting its simpler, shorter temporal structure. This finding aligns with our expectations, given its limited temporal complexity. Overall, these ablation analyses validate the necessity of BCNE's comprehensive temporospatial design for accurately capturing complex, behaviorally relevant dynamics across diverse neural recordings. Importantly, BCNE does not indiscriminately cluster adjacent time points; rather, it represents dynamic variations by drawing genuinely correlated time points closer together while separating adjacent but distinct events within the embedding space.

In BCNE, we employed DNN to unsupervisedly optimize the trajectory generation process, offering key advantages over traditional methods (Figs. 2-5 and Figs. S1, S2, S4). One key benefit is the batch-wise training strategy employed by DNNs, which maintains linear computational complexity, ensuring efficiency and scalability. While similar in principle to parametric t-SNE and parametric UMAP, BCNE significantly outperforms these conventional methods. It achieves this by employing a temporospatial processing integrated with a temporal-aware CNN backbone and recursive latent-manifold refinement, enabling the extraction of intricate temporospatial features that traditional MLP approaches fail to capture. BCNE's scalability is particularly crucial when compared to methods like T-PHATE, which can lead to exponential increases in runtime and memory requirements as the dataset size grows. Consequently, BCNE is highly adaptable to the variable demands of biomedical applications, where dynamic brain data often extends over long periods, ensuring its applicability across a broader range of real-world scenarios. Moreover, we employed a recursive optimization strategy to dive into increasingly deeper latent spaces and subsequently construct manifolds and unfold them into lower-dimensional representations for investigation the cognitive and behavioral process (Fig. 1**b**). As shown in Figs. 3**b** and 4**g**, deepening the recursion depth progressively leads to meaningful clustering for different temporal states, with the most representative performance observed in the Rat hippocampus dataset. In the event segmentation (39-class classification) task of the Sherlock dataset, significant quantitative improvements are observed when using a deeper latent manifold. It is important to note that our reported 2D embedding results serve as a sanity check to assess whether low-dimensional trajectories naturally align with behavioral boundaries, rather than to maximize classification accuracy. In contrast, T-PHATE utilized higher-dimensional embeddings for this experiment in their original paper, which may account for differences in observed outcomes.

In this study, we examine the mechanisms and efficacy of BCNE's recursive optimization strategy for embedding dynamic brain data in an unsupervised manner. We begin by representing each timepoint in the HD dataset using its brain response intensity, then compute two temporal similarity matrices—one in the HD space and one in the embedding space. Minimizing their KL divergence preserves essential information and aligns the output embedding with the original signals. However, since HD data often carry substantial redundancy that diminishes interpretability, we leverage insights from deep learning theory and practice by incorporating progressively deeper latent representations[15–18] into our recursive optimization. Rather than reducing dimensionality in a single shot from the original HD data directly to LD embedding, which generally preserves only local neighborhoods, we propose a multi-stage recursive process where data flow through intermediate dimensionalities defined at different dense layers and is utilized for enhanced brain state trajectory embedding. Empirically, freezing layers 1 through $r$ degraded performance. Allowing these layers to remain adaptive enables the similarity constraints to redistribute representational capacity toward features identified as informative by deeper layers, resulting in improved embedding quality and smoother, more stable convergence. By using these intermediate representations to guide our embedding process, we capture both fine-scale local relationships and broader structural patterns that would be missed in direct dimensionality reduction. In addition to incorporating global information, this sequential dimensionality reduction-based recursive refinement may help to avoid being trapped in local minima, because the multi-stage optimization process provides better initialization for each subsequent step. Since each intermediate representation already approximates the geometry of previous stages, the optimization has a well-informed starting point, thus reducing the risk of getting trapped in poor local minima that plague single-step approaches. Overall, our approach offers a principled alternative to traditional methods that attempt to reduce dimensionality directly from HD to LD space. By compressing information hierarchically, we create embeddings that preserve both local neighborhoods and global topology. This is particularly valuable for neural data, where complex cognitive or behavioral states often manifest as subtle but coherent patterns that span multiple scales of organization. Finally, BCNE offers flexibility in how "smooth" or "segmented" the resulting manifold appears by adjusting recursion depth, as illustrated by BCNE (Recur 0) versus BCNE (Recur 3) in Fig. 3a. Lower recursion depths favor broader, more continuous trajectories, whereas higher depths highlight subtle distinctions and produce event-like clusters. When deeper recursion is used, BCNE inherently yields a series of intermediate embeddings from smooth to more skeletal resolutions, enabling users to select the representation best suited to their analytical objectives.

In discovery-driven brain dynamic data analysis, purely unsupervised/self-supervised methods offer key advantages over supervised learning. Brain data is highly complex, with diverse and often unpredictable patterns. Supervised methods rely on predefined labels, which can limit exploration to known concepts and may overlook novel, emergent relationships. These labels are often human-defined, potentially constraining the discovery of meaningful insights in the data. Unsupervised approaches, on the other hand, allow for open-ended exploration, revealing the intrinsic structure and relationships within the data without relying on prior assumptions. This flexibility is crucial in understanding the complexity of brain function, as it preserves the data's most significant features. Therefore, unsupervised methods are invaluable for uncovering new insights and capturing the essential dynamics of brain activity, offering a more comprehensive understanding of the underlying processes.

The BCNE configurations are thoroughly validated in supplementary (Figs. S3-S12). First, we conducted a comparative analysis between different loss functions, including the proposed manifold learning loss and vanilla contrastive learning loss[27], and demonstrated the advantage of the manifold learning loss in visualizing dynamic brain data. Second, we note that variations in BCNE configuration, including hyperparameters and architecture, lead to alterations in the resulting visualizations[28]. Specifically, the "balance" hyper-parameter, defined as dataset size divided by batch size, tunes the local-to-global view captured by the joint-probability matrices of the HD input data and their LD embeddings. Figs. S7 and S11 show that moderate values (balance equals to 3 or 4) reliably produce more insightful embeddings in our datasets. Finally, we note that our method of correlation projection in BCNE can be further refined. Currently, our approach of computing the temporal affinity matrix neglects long-range temporal correlations to achieve high computational efficiency. If the time interval between recalling the same topic is excessively long and surpasses the predefined drop-off threshold, their correlation might be disregarded by this strategy. Developing an improved correlation analysis method that accounts for these long-range temporal correlations may further enhance the performance of BCNE. Although this study emphasizes LD embeddings (≤3D) to maximize interpretability, BCNE inherently supports arbitrary latent dimensionalities, and systematically exploring higher-dimensional embeddings for predictive tasks represents a compelling direction for future work. Given that BCNE is a newly developed tool for dynamic brain data interpretation, investigating its broader potential applications presents a compelling direction for future research. BCNE may serve as a powerful and intuitive tool for evaluating brain health by comparing trajectory patterns under varying conditions, such as those induced by aging or disease-related neurodegeneration, potentially providing critical insights into the brain's resilience to degradation. Finally, BCNE's adaptability across various brain signal modalities underscores its versatility, extending its utility beyond fundamental neuroscience research into broader domains of computational analysis, clinical decision support, and the interpretation of complex biomedical signals.

In summary, we have developed BCNE, a framework that harnesses temporospatial relationships in dynamic brain data to reveal cognitive and behavioral patterns and visualize brain-state trajectories. Enhanced by recursive deep manifold learning, BCNE significantly improves our ability to uncover diverse temporal states. Experimental studies across various brain datasets highlight BCNE's potential to elucidate intricate dynamic processes. By providing a robust framework for analyzing and interpreting complex brain activity, BCNE advances our understanding of fundamental phenomenon, such as skill acquisition, memory formation, and decision-making, and may inform the study and management of neurological disorders. BCNE's robust analytical capabilities not only support neuroscientists in exploring fundamental brain mechanisms but also hold promise for improving diagnostic accuracy and personalized treatment strategies in clinical neurology. Moreover, it lays the groundwork for developing innovative neurofeedback systems. Furthermore, by enhancing our grasp of fundamental cognitive and behavioral processes, BCNE-driven interpretations of complex brain activities could inform the development of advanced brain-computer interfaces, aid personalized medicine approaches, and guide targeted interventions for neurological disorders.

# Methods

### Dataset

**Sherlock Dataset (Human fMRI BOLD Imaging)**

The Sherlock dataset[20] includes data from sixteen participants who viewed a 48 min clip of the BBC television series 'Sherlock'. Imaging was performed on a Siemens Skyra 3-T scanner with a 20-channel head coil, using a T2*-weighted echo-planar sequence for functional images, and a T1-weighted MPRAGE sequence for anatomical images. The movie was projected using an LCD projector onto a rear-projection screen located in the magnet bore and viewed with an angled mirror. We followed the data preprocessing procedure[13], which includes slice-time and motion correction, linear detrending, high-pass filtering (140s cutoff), and normalization to the Montreal Neurological Institute (MNI) template using FSL. Functional images were resampled to 3mm isotropic voxels, z-scored, and smoothed with a 6mm kernel for analysis. The ROI selection is guided by the references[13,20]. We adopted EV, EA, HV and PMC regions, which are known for their robust response to audiovisual movie stimuli. The voxel-resolution dimensions for the ROIs were: EV = 307, HV = 571, EA = 1,018, PMC = 481.

**Rat Dataset (Hippocampal Electrophysiology)**

The Rat hippocampus dataset[29] detailed electrophysiological recordings from the CA1 hippocampal region of four male Long–Evans rats, obtained using bilaterally implanted silicon probes. Each rat, during the sessions, navigated a 1.6-meter linear track to receive water rewards at both ends. The count of identified putative pyramidal neurons varied from 48 to 120 per rat. We followed the data preprocessing procedure[14], in which neuronal spikes were segmented into 25 ms bins. The spatial position and movement direction (either left or right) of each rat were represented by a 3D vector, incorporating a continuous position value alongside two binary indicators for the direction.

**Macaque Dataset (S1 Electrophysiology)**

The Macaque dataset[24] consists of electrophysiological recordings from Area 2 of the somatosensory cortex (S1) in a rhesus macaque during a center-out reaching task using a manipulandum. In this task, the macaque was required to perform reaches in eight different directions. On half of the trials, the macaque actively executed center-out movements towards a specified target. The other half were 'passive' trials, during which an unexpected 2 Newton force was applied to the manipulandum in one of the eight target directions while the macaque was holding it. The data preprocessing follows reference[14], which followed the procedure outlined by Pei et al.[31], covering a timeframe from −100 ms to 500 ms relative to the onset of movement. The analysis involved segmenting the data into 1 ms time bins and smoothing with a Gaussian kernel (standard deviation = 40 ms).

**Evaluation Metrics**

We applied comprehensive evaluation metrics across three datasets—Sherlock fMRI dataset, rat hippocampal electrophysiological recordings, and macaque S1 electrophysiological data—to assess how effectively our embeddings capture meaningful neurodynamic structures. Formally, let the original brain data be denoted as $\mathbf{X} \in \mathbb{R}^{N \times T}$, where $N$ represents the number of input features (e.g., voxels or electrodes) and $T$ is the number of timepoints. Our method produces a LD embedding $\eta(t) \in \mathbb{R}^O$, with $O$ typically set to 2 or 3 to emphasize brain-state trajectory visualization.

**Sherlock Dataset (Human fMRI BOLD Imaging)**

1) *k-Nearest Neighbors (KNN) Scene Classification*:

We used a k-Nearest Neighbors (KNN) classifier to evaluate how well the embeddings $\eta$ distinguish brain states corresponding to different movie scenes. Each timepoint $t \in \{1, ..., T\}$ has a ground truth category label $L(t) \in \{1, ..., N_{sc}\}$, indicating one of $N_{sc}$ predefined movie scenes. The KNN classifier predicts labels $\hat{L}(t)$ from embeddings $\eta(t)$ across various values of $k$ ($k \in \{1, 3, 5, 8, 10, 30\}$). Classification accuracy is

$$\text{Acc} = \frac{1}{T}\sum_{t=1}^{T} \mathbf{1}(\hat{L}(t) = L(t)), \tag{1}$$

where $\mathbf{1}(\cdot)$ denotes an indicator function:

$$\mathbf{1}\left(\hat{L}(t) = L(t)\right) = \begin{cases} 1, & \text{if } \hat{L}(t) = L(t) \\ 0, & \text{otherwise.} \end{cases} \tag{2}$$

This scene-category KNN provides discrete yet comprehensive assessment, effectively capturing rapid transitions and long-range dependencies in neural activity patterns. Accuracy was determined via 10-fold cross-validation; higher values indicate better separation of scene-specific brain states.

2) *Behaviorally Rated Event Boundaries*:

To evaluate alignment with behaviorally annotated events, we assessed embedding similarities relative to externally defined event boundaries. Each timepoint $t$ has a behavioral event label $L(t)$, partitioning the series into discrete events. Boundaries occur at label transitions ($L(t) \neq L(t-1)$), with series endpoints included. Following references[13,22,23], we constructed paired timepoints at fixed lag $d$, retaining pairs straddling event boundaries. Similarity between embeddings is then:

$$\rho_{\text{event}}(t, t') = \frac{\sum_{i=1}^{O}(\eta_i(t) - \bar{\eta}(t))(\eta_i(t') - \bar{\eta}(t'))}{\sqrt{\sum_{i=1}^{O}(\eta_i(t) - \bar{\eta}(t))^2}\sqrt{\sum_{i=1}^{O}(\eta_i(t') - \bar{\eta}(t'))^2}}, \tag{3}$$

where $\bar{\eta}(t) = \frac{1}{O}\sum_i \eta_i(t)$. Let $\mathcal{W}$ denote the set of all within-event pairs and $\mathcal{B}$ the set of all across-boundary pairs. We then define

$$W_{\text{avg}} = \frac{1}{|\mathcal{W}|}\sum_{(t,t') \in \mathcal{W}} \rho_{\text{event}}(t, t'), \quad B_{\text{avg}} = \frac{1}{|\mathcal{B}|}\sum_{(t,t') \in \mathcal{B}} \rho_{\text{event}}(t, t'). \tag{4}$$

The behavioral difference $D_{\text{event}} = W_{\text{avg}} - B_{\text{avg}}$ quantifies how distinctly the embedding reflects behaviorally defined transitions, with larger values indicating clearer alignment.

**Rat Dataset (Hippocampal Electrophysiology)**

1) *Learning Stage Classification*:

Each timepoint $t$ was labeled with learning stage $L_{\text{stage}}(t)$ based on water-reward thresholds (6, 10, 12, 13, 14, or 18 rewards). To evaluate the robustness of the embeddings in capturing the specific behavioral transitions, we evaluated embedding quality by KNN classification accuracy (as defined previously) for $k \in \{1, 3, 5, 8, 10, 30\}$), using 10-fold cross-validation under each threshold.

2) *Moving Direction Classification*:

We labeled each timepoint by movement direction and computed KNN accuracy on $\eta(t)$ similarly to assess directional encoding in the embedding.

3) *Position–Embedding Representational Similarity*:

This metric evaluates how well the spatial relationships among neural positions are preserved in the embedding. Let $t_i$ and $t_j$ be timepoints with original spatial coordinates $p(t)$ and embeddings $\eta(t)$. Define distance matrices

$$D_p(t_i, t_j) = \|p(t_i) - p(t_j)\|, \tag{5}$$
$$D_\eta(t_i, t_j) = \|\eta(t_i) - \eta(t_j)\|. \tag{6}$$

Extracting the upper-triangle entries ($i < j$), we compute the Pearson correlation

$$\rho_{\text{RSA}} = \frac{\sum_{i<j}(D_p(t_i,t_j) - \bar{D}_p)(D_\eta(t_i,t_j) - \bar{D}_\eta)}{\sqrt{\sum_{i<j}(D_p(t_i,t_j) - \bar{D}_p)^2}\sqrt{\sum_{i<j}(D_\eta(t_i,t_j) - \bar{D}_\eta)^2}}, \tag{7}$$

where $\bar{D}_p$ and $\bar{D}_\eta$ are the means of the respective distance sets. A higher $\rho_{\text{RSA}}$ indicates faithful spatial preservation.

4) *Trustworthiness*:

This metric measures preservation of local neighborhoods in the embedding. For neighborhood size $k$, let $N^{(k)}(t_i)$ be the set of the $k$ nearest neighbors of timepoint $t_i$ in the original space, and let $M^{(k)}(t_i)$ be the corresponding set in the embedding. We define the set of spurious neighbors

$$U_i^{(k)} = M^{(k)}(x_i) \setminus N^{(k)}(x_i). \tag{8}$$

Let $r(t_i, t_j)$ denote the rank of timepoint $t_j$ when all points are sorted by their distance to $t_i$ in the original space. Then

$$T(k) = 1 - \frac{2}{Nk(1N - 3k - 1)} \sum_{i=1}^{N} \sum_{t_j \in U_i^{(k)}} (r(t_i, t_j) - k). \tag{9}$$

Here, $T(k)$ penalizes points that appear close in the embedding but are not truly neighbors in the original space. Higher $T(k)$ values indicate better local structure preservation.

5) *Continuity:*

Continuity quantifies preservation of the original temporal sequence. For each timepoint $t_i$, missing neighbors are $N^{(k)}(t_i) \setminus M^{(k)}(t_i)$. Each such missing neighbor incurs a penalty proportional to its rank in $M^{(k)}(t_i)$. Summing these penalties over all timepoints and normalizing yields the continuity score $C(k) \in [0,1]$, where larger values denote stronger temporal-sequence fidelity.

**Macaque Dataset (S1 Electrophysiology)**

1) *Arm-Angle Classification*:

Each $t$ is labeled $L_{\text{angle}}(t) \in \{0°, 45°, \ldots, 315°\}$. KNN accuracy (as defined previously) on $\eta(t)$ measures angle encoding.

2) *Position–Embedding Representational Similarity*:

For each angle $i$, compute Pearson correlation $\rho_i$ between $D_p$ and $D_\eta$ restricted to timepoints at angle $i$, analogous to the RSA defined above.

**Roll Shift Test**

To assess temporal sensitivity, we circularly shift $p(t)$ by a range of offsets $\delta$ and recompute $\rho_{\text{RSA}}(\delta)$ or $\rho_i$ as above. A peak at $\delta = 0$ with decreasing correlation for $|\delta| > 0$ indicates correct temporal encoding. This test, applied to both rat and macaque data, complements standard RSA to verify spatiotemporal fidelity.

## Temporospatial Correlation-based Reconfiguration

**Temporal Projection**

To effectively visualize the dynamic brain signal with informative temporal trajectory, we employed temporal autocorrelation-based matrix[13] to capture the temporal dynamics of brain activity and reveal the evolving relationships during the dynamic recognition process of a living subject. The brain dynamic signals $X = x_1, x_2, ..., x_T$, where $T$ represents the length of the time series, calculate its autocovariance for each feature $x_t \in \mathbb{R}^N$ separately. The autocorrelation matrix $A \in \mathbb{R}^{N \times (T-1)}$ is calculated as:

$$A(i, \tau) = \frac{1}{T-\tau} \sum_{t=1}^{T-\tau} (X_{t,i} - \bar{X}_i)(X_{t+\tau,i} - \bar{X}_i), \quad (10)$$

where $\tau$ represents the time lag, taking values from 1 to $T$-1 to cover the entire length of the time series. Here $\bar{X}_i$ represents the mean value of the time series for the $i$-th feature, $X_{t,i}$ and $X_{t+\tau,i}$ denote the brain signal intensity at time points $t$ and $t+\tau$, respectively. The autocorrelation matrix is then smoothed using a centered moving average filter:

$$\bar{A}(i, \tau) = \frac{1}{\delta} \sum_{p=\tau-\left[\frac{\delta}{2}\right]}^{\tau+\left[\frac{\delta}{2}\right]} A(i, p), \quad (11)$$

where $\delta$ represents the window size for smoothing. In this process, the convolution operation applies a rolling average, ensuring a centered smoothing effect. If the window exceeds the boundaries of the time series, the effective window size is automatically reduced near the edges, avoiding the need for explicit padding.

We define dropoff point as the first index where the smoothed autocorrelation function becomes negative, serving as a threshold beyond which correlations are considered negligible. Based on the dropoff point $\tau_{drop}$, a time-dynamic correlation matrix $M \in \mathbb{R}^{T \times T}$ is constructed as follows:

$$M(i,j) = \begin{cases} \frac{1}{N} \sum_{f=1}^{N} \bar{A}(f, |i-j|), & \text{if } 0 < |i-j| < \tau_{drop} \\ 0, & \text{otherwise} \end{cases} \quad (12)$$

where $M(i,j)$ represents the strength of correlation between time points $i$ and $j$, incorporating the temporal context captured by the autocorrelation function.

Once the time-dynamic correlation matrix $M$ is obtained, the temporally reconfigured data matrix $\tilde{X} \in \mathbb{R}^{N \times T}$ is computed as:

$$\tilde{X} = XM \oslash [\mathbf{1}_T^\top M], \quad (13)$$

where $\mathbf{1}_T \in \mathbb{R}^T$ is a vector of ones, so that $\mathbf{1}_T^\top M$ produces the row-sums of $M$ as a $1 \times T$ vector, and $\oslash$ denotes elementwise division broadcast across each row of $XM$. Equivalently, the $t$-th column is:

$$\tilde{X}_{,t} = \frac{\sum_{t'=1}^{T} M(t,t') X_{,t'}}{\sum_{t'=1}^{T} M(t,t')}. \quad (14)$$

Here, the new column $\tilde{X}_{,t}$ is a weighted average of the original time-point vectors $X_{,1}, ..., X_{,T}$. A large weight $M(t, t')$ indicates that time points $t$ and $t'$ are strongly correlated, so $X_{,t'}$ contributes more to $\tilde{X}_{,t}$. Dividing by $\sum_{t'=1}^{T} M(t, t')$ ensures that each reconfigured column has a consistent scale, preventing any single time-point cluster from dominating. In essence, Eq.13 smooths and aggregates each timepoint $t$ according to the nearest correlated neighbors in time.

**Image Representation of the Brain Dynamic Data**

*a) Motivation*

We express the responses of the spatial channels (i.e., the HD brain signals) at each time point in terms of a schematically meaningful 2D image representation to effectively incorporate the interactions among recording channels. The transformation maps the responses of the spatial channels to image pixels in a manner that the inter-channel relationships are encoded in the image contextual pattern. Specifically, pairwise inter-channel correlations are computed from the inverse covariance of temporally processed signals, obtained by autocorrelation-based lag-weighted averaging (Fig. 1b). These interactions are encoded contextually into images by minimizing the Gromov-Wasserstein (GW) discrepancy between the interaction matrix and pixelated grid-distance matrix. This process adapts the GenoMap[19] concept, where feature-feature (e.g., gene-gene) interactions are encoded as images for high-performance analysis of HD single-cell RNA-seq data, to dynamic brain signals. By representing inter-channel dependencies in a structured 2D image, standard CNNs can directly exploit these correlations for more effective pattern discernment and downstream analyses.

*b) Quantification of feature–feature interactions*

Let $V = \tilde{X} \in \mathbb{R}^{N \times T}$, the temporally re-weighted matrix from the previous stage, where $N$ is the number of channels and $T$ the number of time points. We estimate the spatial interaction matrix C from the precision matrix $\Omega^{-1}$:

$$C_{ij} = \begin{cases} -\frac{(\Omega^{-1})_{ij}}{\sqrt{(\Omega^{-1})_{ii}(\Omega^{-1})_{jj}}}, & \text{if } i \neq j \\ 1, & \text{if } i = j \end{cases} \quad (15)$$

with the covariance:
$$\Omega_{ij} = \frac{1}{T}\sum_{t=1}^{T}(V_{i,t} - \bar{V}_i)(V_{j,t} - \bar{V}_j), \quad \bar{V}_i = \frac{1}{T}\sum_{t=1}^{T} V_{i,t}. \quad (16)$$
where $V_{i,t}$ is the signal of channel $i$ at time $t$, and $\bar{V}_i$ its temporal mean.

*c) 2D grid template*

We create a square lattice with side $w = \lceil\sqrt{N}\rceil$, assigning each pixel integer coordinates $(x_k, y_k) \in \mathbb{Z}^2$ centered at the origin. The Euclidean distance matrix of the grid $\bar{C} \in \mathbb{R}^{N \times N}$ is
$$\bar{C}_{kl} = \sqrt{(x_k - x_l)^2 + (y_k - y_l)^2}. \quad (17)$$
Both $C$ and $\bar{C}$ are divided by their means so that $\sum_{i,j} C_{ij} = \sum_{i,j} \bar{C}_{ij} = N^2$.

*d) Optimal assignment via GW alignment*

Following optimal-transport (OT) theory, we seek a coupling S that aligns the intra-feature geometry of $C$ with the intra-pixel geometry of $\bar{C}$:
$$d_{GW}(C, \bar{C}, u, v) = \min_{T \in \prod(u,v)} \sum_{i,j,k,l} L(C_{ik}, \bar{C}_{jl}) S_{ij} S_{kl}, \quad (18)$$
where $u = v = \frac{1}{N}\mathbf{1}$ are uniform marginals; $\prod(u,v) = \{S \in \mathbb{R}_+^{N \times N} | S\mathbf{1} = u, S^\top\mathbf{1} = v\}$ is the set of admissible OT plans; $L(a,b) = a \log\frac{a}{b} - a + b$ is the KL divergence loss.

We solve Eq.18 with an entropically-regularized Sinkhorn algorithm[19,32]. Once converged, $S$ is nearly bistochastic; multiplying by $N$ yields a matrix whose rows and columns each sum to 1, effectively acting as a soft permutation from channels to grid locations. A hard assignment is obtained by taking the arg-max in every row, ensuring a one-to-one mapping.

Specifically, $S_{ij}$ represents the probability mass transported from feature $i$ to pixel $j$. Large $S_{ij}$ values indicate that feature $i$ is best placed at grid position j, so channels with strong mutual correlations (large $C_{ik}$) are steered towards spatially proximate pixels (small $\bar{C}_{jl}$).

*e) Generating 2D image representation*

For each time point $t$ the vector $v_t \in \mathbb{R}^N$ is reshaped as
$$\tilde{V}_t = v_t^\top S \in \mathbb{R}^{q \times q}, \quad (19)$$
yielding an image sequence $\tilde{V} \in \mathbb{R}^{T \times q \times q}$. In each image representation $\tilde{V}_t$, interactions between spatial channels are uniquely encoded within the image's contextual pattern, making it well suited for analysis by the subsequent CNN.

It is important to note that the correlations among the spatial channels in our spatial transformation (i.e., generation of a contextual image representation of the dynamic brain data) are directly computed from their mathematical correlations, rather than relying on explicit functional connectivity analyses.

## Technical Framework of BCNE

### Convolutional-based BCNE Architecture

Following the construction of temporospatial correlations, the signal from each subject at each timepoint is reformatted into an image. This allows the temporospatial correlation-induced reconfigured data to be input into the BCNE, with dimensions $N^{ST} \in \mathbb{R}^{q \times s_w \times s_h}$. The network is established based on convolutional architecture, as illustrated in Fig. 1**b**. The architecture initiates with a module that includes four convolutional layers to extract shallow features from the input data in image format. The temporospatial structure of the reconfigured data ensures that points of high correlation are centrally relocated and positioned close to each other, making convolutional layers an apt choice for processing this type of data. They focus on local connections, thus enhancing the depth of the latent space while incurring a significantly lower computational cost compared to that of a purely MLP framework. Following this, a flattening layer converts the 2D latent features into a 1D format. This conversion is succeeded by a fully connected module, consisting of several dense layers. This module projects the data from a global perspective to generate the final output embedding. After the network is trained, the brain signals can directly produce subject-specific visualizations. These visualizations are applicable to the external data embeddings of the same subject's analogous dynamic cognitive and behavioral tasks.

### Recursive Optimization Strategy

*a) Motivation*

An important function of the multiple dense layers in our BCNE is to sequentially reduce the dimensionality of the feature vector representation of the input HD data, from 1,024 in layer 1, 512 in layer 2, 256 in layer 3, 8 in layer 4, and finally to 2 or 3 dimensions. Broadly, each step of the dimensionality reduction here compresses the feature vector to a lower dimension while injecting the vector components with more distant and global relational information of the data as a result of MLP optimization, which captures complex relationships across the data space as information flows through the dense layers[33–36]. Our recursive derivation of the brain state trajectory capitalizes on this unique characteristic of sequential dimensionality reduction to incorporate the information of brain data at different scales. Specifically, in stage 1 (Fig. 1c), we first optimize the BCNE embedding by minimizing its KL divergence with the original HD data. Similar to traditional parametric t-SNE, this embedding primarily takes into consideration the local interactions of the data. However, it benefits from the introduction of image representation and the use of convolutional operations in embedding. In stage 2 (Fig. 1c), the BCNE embedding network is fine-tuned by replacing the original HD input with the feature representation vector from the first dense layer, which helps to integrate the next level of non-local data relationships into the embedding process. The recursion continues to leverage the rich structural information of the subsequent dense layers until the quality of embedding saturates. Empirically, two to three refinement iterations suffice to expose long-range temporal transitions and subtle spatial interactions that a single embedding pass cannot easily uncover. As shown in Fig. 3a–c and Fig. 4b, g, this strategy markedly improves the identification of complex stage transitions and inter-region interactions.

*b) Stage 1: Initial embedding*

BCNE initially trains a deep network using a KL divergence objective similar to parametric t-SNE. This process yields a mapping from the temporospatial preprocessed HD data to an LD space, primarily preserving *local* neighborhoods. Analogous to standard t-SNE, the pairwise similarities ($p_{ij}$, $q_{ij}$) emphasize near neighbors, potentially overlooking broader "global" structures—a particular concern in rich neural data where long-range interactions or transitions can be critical for understanding distinct cognitive or behavioral stages. Specifically, this process involves minimizing the KL divergence between two probability distributions, $P$ and $Q$, where $P$ corresponds to $x$ in the HD space, and $Q$ represents the LD space counterpart, $y$:

$$L_{BCNE} = KL(P||Q) = \sum_{i \neq j} \log \frac{p_{ij}}{q_{ji}} \qquad (20)$$

where $p_{ij}$ and $q_{ij}$ represent normalized pairwise similarities: $p_{ij}$ is the likelihood that a data point $x_i$ in the HD space would select $x_j$ as its neighbor based on the Gaussian distribution. Conversely, $q_{ij}$ is the likelihood that a data point $y_i$ in the LD space would choose $y_j$ as its neighbor based on the t-distribution. The calculation of $p_{ij}$ and $q_{ij}$ is executed as follows:

$$p_{j|i} = \frac{\exp(-\|x_i - x_j\|^2 / 2\sigma_i^2)}{\sum_{k \neq i} \exp(-\|x_i - x_k\|^2 / 2\sigma_i^2)} \qquad (21)$$

$$p_{ij} = \frac{p_{j|i} + p_{i|j}}{2n} \qquad (22)$$

$$q_{ij} = \frac{(1 + \|y_i - y_j\|^2 / \sigma)^{\frac{-\alpha-1}{2}}}{\sum_{k \neq i}(1 + \|y_i - y_j\|^2 / \sigma)^{\frac{-\alpha-1}{2}}} \qquad (23)$$

where $N$ denotes the total number of data points, $\sigma$ represents the Gaussian kernel, and $\alpha$, the degrees of freedom for the t-distribution, is given by $D$-1, where $D$ signifies the desired target dimension.

*c) Stage r ⩾ 1: Progressive global refinement*

  *i) Deeper latent representations*

Following network is trained, each input $x_i$ is passed through the intermediate layers—the deeper part of the BCNE architecture—to extract latent features, denoted by $f^r(x_i)$. These latent representations capture abstract patterns, transforming local elements into more cohesive, higher-level structures.

  *ii) Constructing an Updated HD Similarity*

The latent set $\{f^r(x_i)\}$ is treated as a new "high-dimensional" space from which fresh conditional similarities are computed as:

$$\tilde{p}^r_{j|i} = \frac{\exp(-\|f^r(x_i) - f^r(x_j)\|^2 / 2\sigma_i^2)}{\sum_{k \neq i} \exp(-\|f^r(x_i) - f^r(x_k)\|^2 / 2\sigma_i^2)} \qquad (24)$$

  *iii) Refining the embedding with updated similarities*

We then fine-tune the same BCNE network to match $\tilde{p}_{j|i}$ (derived from deeper latent space) with the LD similarities $q_{j|i}$. This further training optimizes an updated loss

$$L'_{BCNE} = KL(\tilde{P}^r || Q) \qquad (25)$$

Here, $\tilde{P}^r$ is constructed from Eq. 24 and $Q$ is recomputed on the current LD coordinates, adjusting the embedding to concurrently capture local and global structures. At every refinement stage $r$, the network still ingests the original image input $x_i$ at layer 1. Layers $[1, r]$ remain fully trainable and receive gradient updates from $L'_{BCNE}$. Conceptually, each recursive iteration leverages increasingly abstract latent relationships learned by the network. Consequently, the final embedding seamlessly integrates detailed local neighborhoods with overarching manifold topologies.

These three steps are repeated for two to three recursions (e.g., *Dense Layer 1* at *Recur 1* → *Dense Layer 3* at *Recur 3* in our implementation). Each pass "zooms out", weaving the freshly discovered global relationships into the existing manifold while retaining local fidelity. The final embedding unites high-resolution local neighborhoods with coherent long-range topology, exposing subtle state transitions and distributed neural interactions that a single embedding pass fails to reveal. This property is essential for dynamic brain data, where behavioral or cognitive shifts are often encoded in extended, cross-regional patterns.

## Compared Methods

In our evaluation, we considered six methods for comparison: PCA, t-SNE, UMAP, PHATE, T-PHATE, and CEBRA. PCA is a widely adopted dimensionality reduction technique frequently used as a preprocessing step. t-SNE and UMAP are established manifold learning techniques, commonly applied for visualizing HD data. PHATE, a more recent method, leverages Markov diffusion to capture and visualize complex trajectories, particularly in genomic cell differentiation. T-PHATE builds upon PHATE by incorporating atemporal correlations, rendering it particularly well-suited for brain imaging data such as fMRI. CEBRA is a deep learning-based embedding framework specifically designed for brain data analysis, offering multiple operational modes. For the comparison, we employed the CEBRA-Time mode, an unsupervised, data-driven approach that aligns with the goal of BCNE in uncovering intrinsic neural structures without relying on predefined labels. All reported results for CEBRA in this study were obtained using the CEBRA-Time mode.

## Parameter Tuning

In the experiments, we used the default settings for the PCA reimplementation, as the results were not significantly affected. For the other methods, we conducted grid searches to optimize key hyperparameters for each approach.

(i) t-SNE: *Perplexity* $\in \{5, 10, 20, 30, 40, 50\}$, *early exaggeration* $\in \{12, 18, 24, 32\}$.
(ii) UMAP: *Number of neighbors* $\in \{5, 12, 24, 48, 100, 200\}$, *min_dist* $\in \{0.0001, 0.001, 0.01, 0.1, 0.3, 0.5, 0.99\}$.
(iii) PHATE: *n_landmark* $\in \{500, 1000, 2000, N_{temporal\_dim}\}$, $t \in \{1, 3, 5, auto\}$, $knn \in \{5, 10\}$, $decay \in \{20, 40\}$.
(iv) T-PHATE: *n_landmark* $\in \{500, 1000, 2000, N_{temporal\_dim}\}$, $t \in \{1, 3, 5, auto\}$, $knn \in \{5, 10\}$, $decay \in \{20, 40\}$.
(v) CEBRA: *Batch size* $\in \{512, 1024, 2048, 4096\}$, *learning rate* $\in \{1 \times 10^{-4}, 3 \times 10^{-4}\}$, *temperature* $\in \{0.5, 1, 1.5, 2, 3\}$, *distance* $\in \{Cosine, Euclidean\}$.

where $N_{temporal\_dim}$ denotes the temporal dimensionality of the input brain data.

We performed these grid searches on a randomly selected subject from each dataset and then applied the selected hyperparameters to all individual subjects within the same dataset, following the same procedure as the BCNE implementation.

## Final Parameter Selection

We selected the optimal hyperparameters for each method and dataset by maximizing average scores across multiple runs. Below are the final configurations:

(i) Sherlock Dataset
   t-SNE: *perplexity* = 30, *early_exaggeration* = 12.
   UMAP: *min_dist* = 0.1, *n_neighbors* = 5.
   PHATE: *n_landmark* = $N_{temporal\_dim}$, $t$ = 5, $knn$ = 5, *decay* = 20.
   TPHATE: *n_landmark* = $N_{temporal\_dim}$, $t$ = 5, knn = 5, *decay* = 20.
   CEBRA: *batch_size* = 512, *distance* = Euclidean, *learning_rate* = $3 \times 10^{-4}$, *temperature* = 1.0.

(ii) Rat Hippocampus Dataset
   t-SNE: *perplexity* = 10, *early_exaggeration* = 16.
   UMAP: *min_dist* = 0.0001, *n_neighbors* = 24.
   PHATE: *n_landmark* = 2000, $t$ = auto, $knn$ = 10, *decay* = 40.
   TPHATE: *n_landmark* = 2000, $t$ = 1, $knn$ = 10, *decay* = 40.
   CEBRA: *batch_size* = 512, *distance* = Euclidean, *learning_rate* = $1 \times 10^{-4}$, *temperature* = 2.0.

(iii) Macaque Dataset

t-SNE: *perplexity* = 30, *early_exaggeration* = 12.
UMAP: *min_dist* = 0.1, *n_neighbors* = 15.
PHATE: *n_landmark* = 1000, t = 10, *knn* = 10, *decay* = 40.
TPHATE: *n_landmark* = 500, t = 10, *knn* = 5, *decay* = 20.
CEBRA: *batch_size* = 2048, *distance* = Cosine, *learning_rate* = 3 × $10^{-4}$, *temperature* = 1.0.

We observed that PHATE and T-PHATE were the most sensitive to hyperparameter settings across the evaluated datasets, underscoring the importance of meticulous parameter tuning for these methods. In contrast, PCA, t-SNE, UMAP, CEBRA, and BCNE demonstrated relatively stable performance across different parameter configurations. An example of the embeddings generated with varying parameter settings is presented in Fig. S13.

## Data and code availability

1) Sherlock fMRI dataset: https://dataspace.princeton.edu/handle/88435/dsp01nz8062179;
   Preprocessing pipeline: https://github.com/KrishnaswamyLab/TPHATE.
2) Hippocampus dataset: https://crcns.org/data-sets/hc/hc-11/about-hc-11.
3) Macaque dataset: https://datadryad.org/stash/dataset/doi:10.5061/dryad.nk98sf7q7
4) Preprocessing code for hippocampus and macaque datasets: https://github.com/AdaptiveMotorControlLab/CEBRA.
5) BCNE code including the pipeline to replicate all the analyses and preprocessed data can be found in Code Ocean link: https://codeocean.com/capsule/3710904/tree and GitHub: https://github.com/ZixiaZ/BCNE.

## Acknowledgments

This work was partially supported by the National Institutes of Health 1R01CA256890, 1R01CA275772, and 2024 Stanford Human-Centered Artificial Intelligence (HAI) seed grant.

## Author Contributions

ZZ, MT and LX conceived and designed the study; ZZ established the methodology pipeline and did the statistical analyses; YG, QT and YW contributed to refining the methodology; LX implemented quality control of data and the algorithms; ZZ prepared the first draft of the manuscript; ZZ, JL, WW, RF, SL, QW, RY and LX revised the manuscript; All authors contributed to manuscript preparation.

# Ethics declarations

**Competing Interests**

All authors declare no competing interests.

**Animal Data and Ethical Considerations**

All animal data used in this study were obtained from publicly available datasets. This research involved retrospective data analysis, and no new animal experiments were conducted, thus no ethical approval was required.